\documentclass[11pt]{article}

\usepackage{microtype}
\usepackage{cite}
\usepackage{fullpage}
\usepackage{subfigure}
\usepackage{url}
\usepackage{paralist}
\usepackage{enumitem}
\usepackage{graphicx}
\usepackage{multirow}
\usepackage{xspace}
\usepackage{paralist}
\usepackage{color}
\usepackage{mdwlist}
\usepackage{wrapfig}
\usepackage{algpseudocode}
\usepackage{algorithm}
\usepackage{hyperref}
\usepackage[bf]{caption}
\usepackage{boxedminipage}
\usepackage{amssymb,amsmath,amsfonts,amsthm}

\renewcommand{\paragraph}[1]{\vspace{5pt}\noindent\textbf{#1.}}
\newcommand{\ignore}[1]{}

\newcommand{\op}{\mathsf{op}}
\newcommand{\Read}{\mathsf{read}}
\newcommand{\Write}{\mathsf{write}}
\newcommand{\access}{\ensuremath{\mathsf{Access}}\xspace}
\newcommand{\data}{\mathsf{data}}

\newcommand{\position}{\mathsf{position}}

\newcommand{\id}{\ensuremath{\mathsf{a}}\xspace}

\newcommand{\stash}{\ensuremath{S}\xspace}
\newcommand{\ReadBucket}{\ensuremath{\mathsf{ReadBucket}}\xspace}
\newcommand{\WriteBucket}{\ensuremath{\mathsf{WriteBucket}}\xspace}

\newcommand{\name}{Path ORAM\xspace}

\newcommand{\etal}{\emph{et al.}\xspace}

\newcommand{\ceil}[1]{\ensuremath{\left\lceil {#1}\right\rceil}}

\newtheorem{definition}{Definition}

\newtheorem{theorem}{Theorem}
\newtheorem{lemma}{Lemma}

\newtheorem{proposition}{Proposition}

\newcommand{\Path}{\mathcal{P}}

\renewcommand{\vector}[1]{\mathbf{#1}}

\begin{document}

\title{{\LARGE \bf Path ORAM:\\ An Extremely Simple Oblivious RAM Protocol}}

\author{
Emil Stefanov$^\dagger$,
Marten van Dijk$^\ddagger$,
Elaine Shi$^\ast$,
T-H. Hubert Chan$^{**}$,
Christopher Fletcher$^\circ$,\\
Ling Ren$^\circ$,
Xiangyao Yu$^\circ$,
Srinivas Devadas$^\circ$
\\
\\
$^\dagger$UC Berkeley\hspace{1cm}
$^\ddagger$UConn\hspace{1cm}
$^\ast$UMD\hspace{1cm}
$^{**}$University of Hong Kong\hspace{1cm}
$^\circ$MIT CSAIL
}

\date{\large Our algorithm first appeared on February 23, 2012 on arXiv.org.\\
\url{http://arxiv.org/abs/1202.5150v1}}

\maketitle
\begin{abstract}
We present Path ORAM, an extremely simple Oblivious RAM protocol with a small amount of client storage. Partly due to its simplicity, Path ORAM is the most practical ORAM scheme known to date with small client storage. We formally prove that Path ORAM has a $O(\log N)$ bandwidth cost for blocks of size $B = \Omega(\log^2 N)$ bits. For such block sizes, Path ORAM is asymptotically better than the best known ORAM schemes with small client storage. Due to its practicality, Path ORAM has been adopted in the design of secure processors since its proposal.
\end{abstract}

\section{Introduction}

It is well-known that data encryption alone is often not enough to protect users' privacy 
in outsourced storage applications. 
The sequence of storage locations accessed by the client (i.e., access pattern) can leak a significant amount of sensitive information about the unencrypted data through statistical inference. For example, Islam et. al. demonstrated that by observing accesses to an encrypted email repository, an adversary can infer as much as 80\% of the search 
queries~\cite{accesspatternleak}.

Oblivious RAM (ORAM) algorithms, first proposed by Goldreich and Ostrovsky~\cite{GSORAM}, allow a client to conceal its access pattern to the remote storage 
by continuously shuffling and re-encrypting data as they are accessed. 
An adversary can observe the physical storage locations accessed, 
but the ORAM algorithm ensures that the adversary has negligible 
probability of learning anything about the true (logical) access pattern.
Since its proposal, the research community has strived
to find an ORAM scheme that is not only theoretically  
interesting, but also practical~\cite{oram01,oram02,oram03,GMOT11,oram05,oram06,oram07,oram08,oram09,GoldORAM,OsORAM,oram12,oram13,oram14,ndss12,asiacrypt11,oblivistore,shroud,privatefs}.

In this paper, we propose 
a novel ORAM algorithm called 
\textit{Path ORAM}\footnote{Our construction is called Path ORAM because data on the server is always accessed in the form of tree paths.}.
This is to date the most practical ORAM construction under small
 client storage. 
We prove theoretical bounds on its performance and also present matching
experimental results.

Path ORAM makes the following contributions: 

\begin{table*}[t]
\centering
{
\begin{tabular}{|c|c|c|c|c|c|}
\hline
\multirow{2}{*}{ORAM Scheme} & {Client Storage} & {Read \& Write Bandwidth} \\ 
& (\# blocks of size $B$) & (\# blocks of size $B$)  \\ \hline
&  & \\[-5pt]
{Kushilevitz \etal\cite{oram03} ($B = \Omega(\log N)$)} & {$O(1)$} & {$O(\log^2 N / \log \log N)$}  \\[2pt]
\hline
 & & \\[-5pt]
Gentry \etal~\cite{gentryoram} ($B = \Omega(\log N)$) & 
$O(\log^2 N)\cdot\omega(1)$  & 
$O(\log^3 N/\log\log N)\cdot\omega(1)$  \\
\hline
 & & \\[-5pt]
Chung \etal~\cite{chungoram2} ($B = \Omega(\log N)$) & 
\multirow{2}{*}{$O(\log^{2+\epsilon}(N))$}  & 
\multirow{2}{*}{$O(\log^2 N \cdot\log\log N)\cdot\omega(1)$} \\
(concurrent work) & &
\\[2pt]\hline
\hline
 & &\\[-5pt]
Recursive \name & \multirow{2}{*}{$O(\log N)\cdot\omega(1)$} & \multirow{2}{*}{$O(\log^2 N)$} \\[2pt]
for small blocks, $B = \Omega(\log N)$ &  & \\
\hline
 & &\\[-5pt]
{Recursive \name} & \multirow{3}{*}{$O(\log N)\cdot\omega(1)$} & \multirow{3}{*}{$O(\log N) $} \\
for moderately sized blocks, $B = \Omega(\log^2 N)$ &  & \\
(we use a block size of $\Theta(\log N)$ during recursion)&  & \\
\hline
\end{tabular}}
\caption{{\bf Comparison to other ORAM schemes.}
\normalfont
$N$ is the total number of blocks and $B$ is the block size (in bits).
The failure probability is set to $N^{-\omega(1)}$ in this table, i.e.,
negligible in $N$.
This table assumes that the stashes are stored on the client side.  
}
\label{tab:ORamComparison}
\end{table*}

\paragraph{Simplicity and practical efficiency}
In comparison to other ORAM algorithms, our construction is arguably much simpler. Although we have no formal way of measuring its simplicity, the core of the \name algorithm can be described in just \ref{line:AccessLineCount} lines of pseudocode (see Figure~\ref{fig:Access}) and our construction does not require performing sophisticated deamortized oblivious sorting and oblivious cuckoo hash table construction like many existing 
ORAM algorithms~\cite{GSORAM,oram01,oram02,oram03,GMOT11,oram05,oram06,oram07,oram08,oram09,GoldORAM,OsORAM,oram12,oram13,oram14}.
Instead, each ORAM access can be expressed as simply fetching and storing a single path in a tree stored remotely on the server. 
Path ORAM's simplicity
makes it more practical  than any existing
ORAM construction with small (i.e., constant or poly-logarithmic) local storage. 

\paragraph{Asymptotic efficiency}
We prove that for a reasonably large 
block size $B = \Omega(\log^2 N)$ bits where $N$ is the total number of 
blocks (e.g., 4KB blocks), recursive Path ORAM achieves an asymptotic 
\textit{bandwidth cost} of $O(\log N)$ blocks, and consumes
$O(\log N)\omega(1)$ blocks of client-side storage\footnote{
Throughout this paper, when we write the notation $g(n) = O(f(n)) \cdot \omega(1)$, we mean that  for any function $h(n) = \omega(1)$, it holds that $g(n) = O( f(n) h(n) )$. Unless otherwise stated, all logarithms are assumed to be in base 2.}. In other words, to access a single logical block, the client needs to access $O(\log N)$ physical blocks to hide its access patterns from the storage server. The above result achieves a failure probability of 
$N^{-\omega(1)}$, negligible in $N$.

As pointed out later in Section~\ref{sec:related}, our result outperforms the best known ORAM for small client storage~\cite{oram03}, both in terms of asymptotics and practicality, for reasonably large block sizes, i.e., block sizes typically encountered in practical applications.

\paragraph{Practical and theoretic impact of Path ORAM}
Since we first proposed Path ORAM~\cite{OriginalPathORam} in February 2012, it has made both a practical and a theoretic impact in the community. 

On the practical side, Path ORAM is the most suitable known algorithm for hardware
ORAM implementations due to its conceptual simplicity,
small client storage,
and practical efficiency.
Ren \etal built a simulator for an ORAM-enabled secure processor based on the Path ORAM algorithm~\cite{REN13} and the Ascend processor architecture~\cite{ascend-stc12,cwf-masters} uses Path ORAM as a primitive.
Maas \etal~\cite{phantom}
implemented Path ORAM on a secure processor using FPGAs
and the Convey platform.

On the theoretic side,  
subsequent to the proposal of Path ORAM, several theoretic works adopted the same idea of 
path eviction in their ORAM constructions
--- notably the works by Gentry \etal~\cite{gentryoram} 
and Chung \etal~\cite{chungoram,chungoram2}.
These two works also try to improve ORAM bounds  
based on the binary tree construction by Shi \etal~\cite{asiacrypt11};
however, as pointed out in Section~\ref{sec:related}
our bound is asymptotically better than those by Gentry \etal~\cite{gentryoram} and  
Chung \etal~\cite{chungoram,chungoram2}. 
Gentry's Path ORAM variant 
construction has also been applied to secure multiparty computation~\cite{gentryoram}.

\paragraph{Novel proof techniques}
Although our construction is simple, the proof for upper bounding the client storage is quite intricate and interesting. 
Our proof relies on an abstract infinite ORAM construction used only for analyzing
the stash usage of a non-recursive Path ORAM.
For a non-recursive Path ORAM,
we show that during a particular operation, the probability that
the stash stores more than $R$ blocks is at most $\leq 14\cdot 0.6002^{-R}$.
For certain choices of parameters (including $R$) and $N$ data blocks, 
our recursive Path ORAM construction has at most $\log N$ levels,
the stash at the client storage has a capacity of $R \log N$ blocks, the server
storage is $20N$ blocks, and the bandwidth is $10(\log N)^2$ blocks per load/store operation.
For $s$ load/store operations, a simple union bound can show that this recursive Path ORAM
fails during one of the $s$ load/store operations
(due to exceeding stash capacity) with probability at most
$\leq 14 s \log N \cdot 0.625^{-R}$.  Choosing $R = \Theta(\log s + \log \log N) \cdot \omega(1)$ can make the failure probability negligible.  We shall refine
the parameters and do a more careful analysis to 
achieve $\Theta(\log N) \cdot \omega(1)$ blocks of client storage in the
recursive construction.
Our empirical results in Section~\ref{sec:experiments} indicate that the constants in practice are even lower than our theoretic bounds.

\subsection{Related Work}
\label{sec:related}
Oblivious RAM was first investigated by Goldreich and Ostrovsky~\cite{OsORAM,GoldORAM,GSORAM} in the context of protecting software from piracy, and efficient simulation of programs on oblivious RAMs. Since then, there has been 
much subsequent work \cite{GSORAM,oram01,oram02,oram03,GMOT11,oram05,oram06,oram07,oram08,oram09,GoldORAM,OsORAM,oram12,gentryoram,chungoram} devoted to improving ORAM constructions.
Path ORAM is based upon the binary-tree ORAM framework proposed
by Shi \etal~\cite{asiacrypt11}.

\paragraph{Near optimality of Path ORAM}
Under small (i.e., constant or poly-logarithmic) client storage,
the best known ORAM was proposed by Kushilevitz \etal, and 
has $O(\log^2 N/\log \log N)$ blocks
bandwidth cost~\cite{oram03}.

Path ORAM achieves an asymptotic improvement under
reasonable assumptions about the block size: 
when the block size is 
at least 
$\Omega(\log^2 N)$ bits, and
using a smaller block size of $\Theta(\log N)$ for the position map levels,  
Path ORAM has bandwidth cost of $O(\log N)$.
Such a block size $\Omega(\log^2 N)$ bits  
is what one  typically encounters 
in most practical applications (e.g., 4KB blocks in file systems).
Goldreich and Ostrovsky show that under $O(1)$ client
storage, any ORAM algorithm must have bandwidth cost
$\Omega(\log N)$ (regardless of the block size).
Since then, a long-standing open question 
is whether it is possible to have an ORAM construction that
has $O(1)$ or $poly \log (N)$
client-side storage and $O(\log N)$ blocks bandwidth cost~\cite{GSORAM,oram09,oram03}.
Our bound partially addresses this open question for reasonably large block sizes.

\paragraph{Comparison with Gentry \etal and Chung \etal}
Gentry \etal~\cite{gentryoram} improve on the binary tree ORAM scheme
proposed by Shi \etal~\cite{asiacrypt11}.
To achieve $2^{-\lambda}$ failure probability, their
scheme achieves 
$O(\lambda (\log N)^2/ (\log \lambda))$ blocks bandwidth cost,
for block size 
$B = \log N$ bits.
Assuming that $N = poly(\lambda)$,
their bandwidth cost is 
$O(\lambda \log N)$ blocks. 
In comparison, recursive Path ORAM achieves 
$O(\log^2 N)$ blocks
bandwidth cost when $B = \log N$.
Note that typically $\lambda \gg \log N$ since $N = poly (\lambda)$.
Therefore, recursive Path ORAM is much more efficient than the scheme by Gentry \etal.  
Table~\ref{tab:ORamComparison}
presents this comparison, assuming a failure probability
of $N^{-\omega(1)}$, i.e., negligible in $N$.
Since $N = poly(\lambda)$, the failure probability can
also equivalently be written as $\lambda^{-\omega(1)}$. 
We choose to use $N^{-\omega(1)}$
to simplify the notation in the asymptotic bounds.

Chung and Pass~\cite{chungoram} 
proved a similar (in fact slightly worse) bound as 
Gentry \etal~\cite{gentryoram}. As mentioned earlier, 
our bound  	
is asymptotically better than Gentry \etal~\cite{gentryoram}
or Chung and Pass~\cite{chungoram}.

In recent concurrent and independent work, 
Chung \etal proposed another statistically secure binary-tree ORAM algorithm~\cite{chungoram2} based on Path ORAM. 
Their theoretical bandwidth bound is a $\log \log n$ factor worse than ours for blocks of size $\Omega(\log N)$.
Their simulation results suggest an empirical bucket size of $4$~\cite{personalchung} --- which means that their
practical bandwidth cost is a constant factor worse than Path ORAM, since they require operating on $3$ paths  
in expectation for each data access, while Path ORAM requires reading and writing only $1$ path.

\paragraph{Statistical security}
We note that Path ORAM is also statistically secure (not counting the encryption).
Statistically secure ORAMs have been studied
in several prior works~\cite{perfectoram,perfectoram00}.
All known binary-tree based ORAM schemes and variants
are also statistically secure~\cite{asiacrypt11,chungoram,gentryoram} 
(assuming each bucket is a trivial ORAM).

\section{Problem Definition}

We consider a client that wishes to store data at a remote untrusted server while preserving its privacy. While traditional encryption schemes can provide data confidentiality, they do not hide the data access pattern which can reveal very sensitive information to the untrusted server. In other words, the blocks accessed on the server and the order in which they were accessed is revealed. We assume that the server is untrusted, and the client is trusted, including the client's 
processor, memory, and disk.

The goal of ORAM is to completely hide the data access pattern (which blocks were read/written) from the server. From the server's perspective, the data access patterns from two
sequences of read/write operations with the same length must be indistinguishable.

\paragraph{Notations}
We assume that the client fetches/stores data on the server in atomic units, referred to as \textit{blocks}, of size $B$ bits each. For example, a typical value for $B$ for cloud storage is $64-256$ KB while for secure processors smaller blocks ($128$~B to $4$~KB) are preferable. Throughout the paper, let $N$ be the working set, i.e., the number of distinct data blocks that are stored in ORAM.

\paragraph{Simplicity}
We aim to provide an extremely simple ORAM construction in contrast with previous work. Our scheme consists of only \ref{line:AccessLastLine} lines of pseudo-code as shown in Figure~\ref{fig:Access}.

\paragraph{Security definitions}
We adopt the standard security definition for ORAMs from \cite{ndss12}.
Intuitively, the security definition requires that the server
learns nothing about the access pattern. In other words,
no information should be leaked about:
1) which data is being accessed;
2) how old it is (when it was last accessed);
3) whether the same data is being accessed (linkability);
4) access pattern (sequential, random, etc);
or 5) whether the access is a read or a write.
\begin{definition}[Security definition]
Let $$\vec{y} := ((\op_M, \id_M, \mathsf{data}_M), 
\ldots, 
(\op_1, \id_1, \mathsf{data}_1))$$
denote a data request sequence of length $M$, 
where each $\op_i$ denotes a 
$\Read(\id_i)$ or a $\Write(\id_i, \mathsf{data})$ operation.
Specifically, $\id_i$ denotes the identifier of the block being read or written, 
and $\mathsf{data}_i$ denotes the data being written.
In our notation, index $1$ corresponds to the most recent load/store and index $M$ corresponds to the oldest load/store operation.

Let $A(\vec{y})$ denote the (possibly randomized) sequence of 
accesses to the remote storage
given the sequence of data requests $\vec{y}$.
An ORAM construction is said to be secure
if (1) for any two data request sequences $\vec{y}$ and $\vec{z}$ of the same length, 
their access patterns 
$A(\vec{y})$ and $A(\vec{z})$ are computationally indistinguishable
by anyone but the client, and (2) the ORAM construction is correct in the sense that it returns on input $\vec{y}$ data that is consistent with $\vec{y}$ with probability $\geq 1-negl(|\vec{y}|)$, i.e., the ORAM may fail with probability $negl(|\vec{y}|)$.
\label{def:Obliviousness}
\end{definition}

Like all other related work,  our ORAM constructions do not consider information leakage through the timing channel, such as when or how frequently the client makes data requests.
Achieving integrity 
against a potentially malicious server is discussed in Section~\ref{sec:integrity}.
We do not focus on integrity in our main presentation. 

\section{The Path ORAM Protocol}
\label{sec:protocol}

We first describe the Path ORAM protocol with linear amount of client storage, and then later in Section~\ref{sec:recursion} we explain how the client storage can be reduced to 
(poly-)logarithmic via recursion. 

\subsection{Overview}
We now give an informal overview of the Path ORAM protocol. The client stores a small amount of local data in a stash. The server-side storage is treated as a binary tree where each node is a bucket that can hold up to a fixed number of blocks.

\paragraph{Main invariant}
We maintain the invariant that at any time, each block is mapped to a uniformly random leaf bucket in the tree, and unstashed blocks are always placed in some bucket along the path to the mapped leaf.

Whenever a block is read from the server, the entire path to the mapped leaf is read into the stash, the requested block is remapped to another leaf, and then the path that was just read is written back to the server. When the path is written back to the server, additional blocks in the stash may be evicted into the path as long as the invariant is preserved and there is remaining space in the buckets.

\subsection{Server Storage}

Data on the server is stored in a tree consisting of buckets as nodes. The tree does not have to necessarily be a binary tree, but we use a binary tree in our description for simplicity.

\paragraph{Binary tree} The server stores a binary tree data structure of height $L$ and $2^L$ leaves. In our theoretic bounds, we need $L=\ceil{\log_2(N)}$, but in our experiments,
we observe that $L=\ceil{\log_2(N)}-1$ is sufficient.  The tree can easily be laid out as a flat array when stored on disk. The levels of the tree are numbered $0$ to $L$ where level $0$ denotes the root of the tree and level $L$ denotes the leaves.

\paragraph{Bucket} Each node in the tree is called a bucket. Each bucket can contain up to $Z$ real blocks. If a bucket has less than $Z$ real blocks, it is padded with dummy blocks to always be of size $Z$. It suffices to choose the bucket size $Z$ to be a small constant such as $Z=4$ (see Section~\ref{sec:StashDistribution}).

\paragraph{Path}
Let $x \in \{0, 1, \ldots, 2^L -1 \}$ 
denote the $x$-th leaf node in the tree.
Any leaf node $x$ defines a unique path from leaf $x$ to the root of the tree. We use $\Path(x)$ to denote set of buckets along the path from leaf $x$ to the root.   
Additionally, $\Path(x, \ell)$ denotes the bucket in $\Path(x)$ at level $\ell$ in the tree. 

\paragraph{Server storage size}
Since there are about $N$ buckets in the tree, the total server storage used is about $Z \cdot N$ blocks.

\subsection{Client Storage and Bandwidth}
The storage on the client consists of $2$ data structures, a stash and a position map:

\paragraph{Stash} 
During the course of the algorithm, a small number of blocks might overflow
from the tree buckets on the server.
The client locally stores these overflowing
blocks in a 
local data structure \stash called the stash. In Section~\ref{sec:fail}, we prove that the stash has a worst-case size of $O(\log N) \cdot \omega(1)$ blocks with high probability. In fact, in Section~\ref{sec:BucketLoad}, we show that the stash is usually empty after each ORAM read/write operation completes.

\paragraph{Position map}
The client stores a position map, such that $x := \position[\id]$ 
means that block $\id$ is currently mapped 
to the $x$-th leaf node --- this means that block $\id$ 
resides in some bucket in path $\Path(x)$, or in the stash. The position map changes over time as blocks are accessed and remapped.

\paragraph{Bandwidth}
For each load or store operation, the client reads a path of $Z \log N$ blocks from the server and then writes them back, resulting in a total of $2Z \log N$ blocks bandwidth used per access. Since $Z$ is a constant, the bandwidth usage is $O(\log(N))$ blocks.

\paragraph{Client storage size}
For now, we assume that the 
position map and the stash are both stored
on the client side.
The position map is of size $NL=N\log N$ bits, which is of size $O(N)$ blocks when the block size $B=\Omega(\log N)$.
In Section~\ref{sec:fail}, we prove that the stash for the basic non-recursive Path ORAM is at most $O(\log N)\omega(1)$ blocks to obtain 
negligible failure probability. 
Later in in Section~\ref{sec:recursion}, we explain how the recursive
construction can also achieve client storage of $O(\log N) \cdot \omega(1)$ blocks as shown in Table~\ref{tab:ORamComparison}.

{\footnotesize
\begin{table}[t]
\centering
{
\begin{tabular}{|c|p{0.65\columnwidth}|}
\hline
$N$ & Total \# blocks outsourced to server\\ \hline 
$L \ignore{= \lceil \log_2 N \rceil}$ & Height of binary tree \\ \hline 
$B$ & Block size (in bits) \\ \hline 
$Z$ & Capacity of each bucket (in blocks)\\ \hline 
$\Path(x)$ & path from leaf node $x$ to the root\\ \hline
$\Path(x, \ell)$ & the bucket at level $\ell$ along the path $\Path(x)$\\ \hline
$\stash$ & client's local stash \\ \hline
$\position$ & client's local position map\\ \hline
$x := \position[\id]$ & block $\id$ is currently associated with leaf node $x$, i.e.,
block $\id$ resides somewhere along $\Path(x)$ or in the stash.\\\hline
\end{tabular}
\caption{Notations}
}
\end{table}
}

\subsection{Path ORAM Initialization}
\label{sec:initialize}
The client stash \stash is initially empty. 
The server buckets are intialized 
to contain random encryptions of the dummy block (i.e., initially
no block is stored on the server).
The client's position map is filled with independent 
random numbers between $0$ and  $2^L -1$.

\subsection{Path ORAM Reads and Writes}

\begin{figure}[t]
\centering
\begin{boxedminipage}{\columnwidth}
\underline{$\access(\op, \id, \data^*)$}: 
\begin{algorithmic}[1]

\vspace{0.2cm}
\State $x \leftarrow \position[\id]$
\label{line:GetPosition}
\label{line:RemapBlockStart}
\State $\position[\id] \leftarrow \mathsf{UniformRandom}(0 \ldots 2^L-1)$
\label{line:SetPosition}
\label{line:RemapBlockEnd}

\vspace{0.2cm}
\For{$\ell \in \{0, 1, \ldots, L\}$} \label{line:ReadPathStart}
	\State $\stash \leftarrow \stash \cup \mathsf{ReadBucket}(\Path(x, \ell))$
\EndFor \label{line:ReadPathEnd}

\vspace{0.2cm}
\State $\data \leftarrow$ Read block \id from $\stash$ 
\label{line:UpdateBlockStart}
\If{$\op = \Write$}
	\State $\stash \leftarrow (\stash - \{(\id, \data)\}) \cup \{(\id, \data^*)\}$
\EndIf \label{line:UpdateBlockEnd}
\vspace{0.2cm}
\For{$\ell \in \{L,L-1,\ldots,0\}$} \label{line:WritePathStart}
	\State $S' \leftarrow \{ (\id', \data') \in \stash : \Path(x, \ell) = \Path(\position[\id'], \ell) \}$
	\label{line:FindWritableBlocks}
	\State $S' \leftarrow$ Select $\min(|S'|, Z)$ blocks from $S'$.
	\State $\stash \leftarrow \stash - S'$
	\State $\mathsf{WriteBucket}(\Path(x, \ell), S')$
\EndFor \label{line:WritePathEnd}

\vspace{0.2cm}
\State \Return $\mathsf{\data}$ \label{line:AccessLineCount}

\label{line:AccessLastLine}
\end{algorithmic}
\end{boxedminipage}
\caption{
{\bf Protocol for data access}.
\normalfont Read or write a data block identified by $\id$. If $\op = \Read$, the input parameter $\data^* = \mathsf{None}$, and the \access operation reads block \id from the ORAM. If $\op = \Write$, the \access operation writes the specified $\data^*$ to the block identified by \id and returns the block's old data.
}
\label{fig:Access}
\end{figure}

In our construction, reading and writing a block to ORAM is done via a single protocol called \access described in Figure~\ref{fig:Access}. Specifically, to read block \id, the client performs
$\mathsf{data} \leftarrow \access(\Read, \id, \mathsf{None})$
and to write $\mathsf{data}^*$ to block \id, the client performs
$\access(\Write, \id, \mathsf{data}^*)$.
The \access protocol can be summarized in 4 simple steps:

\begin{enumerate*}
	\item \textbf{Remap block} (Lines \ref{line:RemapBlockStart} to \ref{line:RemapBlockEnd}): Randomly remap the position of block $\id$ to a new random position. Let $x$ denote the block's old position.
	\item \textbf{Read path} (Lines \ref{line:ReadPathStart} to \ref{line:ReadPathEnd}): Read the path $\Path(x)$ containing block $\id$.
	\item \textbf{Update block} (Lines \ref{line:UpdateBlockStart} to \ref{line:UpdateBlockEnd}): If the access is a write, update the data stored for block $\id$.
	\item \textbf{Write path} (Lines \ref{line:WritePathStart} to \ref{line:WritePathEnd}): Write the path back and possibly include some additional blocks from the stash if they can be placed into the path. Buckets are greedily filled with blocks in the stash in the order of leaf to root, ensuring that blocks get pushed as deep down into the tree as possible. A block $\id'$ can be placed in the bucket at level $\ell$ only if the path $\Path(\position[\id'])$ to the leaf of block $\id'$ intersects the path accessed $\Path(x)$ at level $\ell$. In other words, if $\Path(x, \ell) = \Path(\position[\id'], \ell)$.
\end{enumerate*}

Note that when the client performs \access on a block for the first time, it will not find it in the tree or stash, and should assume that the block has a default value of zero.

\paragraph{Subroutines}
We now explain the $\ReadBucket$ and the $\WriteBucket$ subroutine.
For $\ReadBucket(\mathsf{bucket})$, the client reads all $Z$ blocks 
(including any dummy blocks) from the bucket stored on the server. 
Blocks are decrypted as they are read.

For $\WriteBucket(\mathsf{bucket}, \mathsf{blocks})$,
the client writes the blocks $\mathsf{blocks}$ into the  
specified $\mathsf{bucket}$
on the server. When writing, the client pads $\mathsf{blocks}$ with dummy blocks to 
make it of size $Z$ --- note that this is important for security.
All blocks (including dummy blocks) are re-encrypted, using a randomized encryption scheme, 
as they are written.

\paragraph{Computation}
Client's computation is $O(\log N) \cdot \omega(1)$ per data access.
In practice, the majority of this time is spent 
decrypting and encrypting $O(\log N)$ blocks per data access. We treat the server as a network storage device, so it only needs to do the computation necessary to retrieve and store $O(\log N)$ blocks per data access.

\subsection{Security Analysis} To prove the security of Path-ORAM, let $\vec{y}$ be a data request sequence of size $M$. By the definition of Path-ORAM, the server sees $A(\vec{y})$ which is a sequence
$$\mbox{{\bf p}}=(\position_M[\id_M],\position_{M-1}[\id_{M-1}], \ldots, \position_1[\id_1]),$$
where $\position_j[\id_j]$ is the position of address $\id_j$ indicated by the position map for the $j$-th load/store operation,
together with a sequence of encrypted paths ${\cal P}(\position_j(\id_j))$, $1\leq j\leq M$, each encrypted using randomized encryption. The sequence of encrypted paths is computationally indistinguishable from a random sequence of bit strings by the definition of randomized encryption (note that ciphertexts that correspond to the same plaintext use different randomness and are therefore indistinguishable from one another). 
The order of accesses from $M$ to 1 follows the notation from Definition~\ref{def:Obliviousness}.

Notice that once $\position_i(\id_i)$ is revealed to the server, it is remapped to a completely new random label, hence, $\position_i(\id_i)$ is statistically independent of $\position_j(\id_j)$ for $j<i$ with $\id_j=\id_i$. Since the positions of different addresses do not affect one another in Path ORAM, $\position_i(\id_i)$ is statistically independent of $\position_j(\id_j)$ for $j<i$ with $\id_j\neq \id_i$. This shows that $\position_i(\id_i)$ is statistically independent of $\position_j(\id_j)$ for $j<i$, therefore, (by using Bayes rule)  $\Pr(\mbox{{\bf p}}) = \prod_{j=1}^M \Pr(\position_j(\id_j)) = (\frac{1}{2^L})^M$. This proves that $A(\vec{y})$ is computationally indistinguishable from a random sequence of bit strings.

Now the security follows from Theorem~\ref{thm:main} in Section~\ref{sec:fail}: For a stash size $O(\log N) \cdot \omega(1)$ Path ORAM  fails (in that it exceeds the stash size) with at most negligible probability.

\section{Recursion and Parameterization}
\label{sec:recursion}

\subsection{Recursion Technique}

In our non-recursive scheme described in the previous section,
the client must store a relatively large 
position map.
We can leverage the same recursion idea as described in the ORAM constructions of Stefanov \etal~\cite{ndss12} and Shi \etal~\cite{asiacrypt11}.
to reduce the client-side storage. 
The idea is simple: instead of storing the position map on the client
side, we store the position map on the server side in a smaller ORAM, and recurse.

More concretely, consider a recursive Path ORAM made up of a series of ORAMs called $\mathsf{ORam}_0,\mathsf{ORam}_1,\mathsf{ORam}_2, \dots, \mathsf{ORam}_X$ where $\mathsf{ORam}_0$ contains the data blocks, the position map of $\mathsf{ORam}_i$ is stored in $\mathsf{ORam}_{i+1} $, and the client stores the position map for $\mathsf{ORam}_X$.
To access a block in $\mathsf{ORam}_0$, the client looks up its position in $\mathsf{ORam}_1$, which triggers a recursive call to look up the position of the position in $\mathsf{ORam}_2$, and so on until finally a position of $\mathsf{ORam}_X$ is looked up in the client storage.
For a more detailed description of the recursion technique,
we refer the readers to~\cite{ndss12,asiacrypt11}.

\subsection{Parameterization}

We can choose the block size for the recursive ORAMs to parametrize the recursion.

\paragraph{Uniform Block Size}
Suppose each block has size $\chi \log N$ bits, where $\chi \geq 2$.
This is a reasonable assumption that has been made by Stefanov \etal~\cite{ndss12,oblivistore} and Shi \etal~\cite{asiacrypt11}. For example, a standard 4KB block consists of $32768$ bits and this assumption holds for all $N \le 2^{16382}$. 
In this case,
the number of level of recursions is $O(\frac{\log N}{\log \chi})$.  Hence, the
bandwidth overhead is $O(\frac{\log^2 N}{\log \chi})$ blocks.

\begin{compactitem}
\item[(i)] \textbf{Separate Local Storage for Stashes from Different Levels of Recursion.} 
From Theorem~\ref{thm:main} in Section~\ref{sec:fail},
in order to make the failure probability for each stash to be $\frac{1}{N^{\omega(1)}}$,
the capacity of each stash can be set to $\Theta(\log N) \cdot \omega(1)$ blocks.
Hence, storing stashes from all levels needs $O(\frac{\log^2 N}{\log \chi}) \cdot \omega(1)$ blocks
of storage.
\item[(ii)] \textbf{Common Local Storage for Stashes from All Levels of Recursion.} 
A common local storage can be used to store stash blocks from all levels of recursion.
From Theorem~\ref{thm:main}, it follows that the number of blocks in each stash
is dominated by some geometric random variable.  Hence, it suffices to analyze
the sum of $O(\frac{\log N}{\log \chi})$ independent geometric random variables.
From the analysis in Section~\ref{sec:common_stash} (with the detailed proof
given in Section~\ref{sec:SharedStashBounds}), it follows that the capacity of the common storage
can be set to $\Theta(\log N)\cdot \omega(1)$ to achieve failure probability 
$\frac{1}{N^{\omega(1)}}$.
\end{compactitem}

\paragraph{Non-uniform Block Size}
Suppose that the \textit{regular block} size is $B = \Omega(\log^2 N)$ bits. This is the block size for the original ORAM ($\mathsf{ORam}_0$). Suppose for the position map ORAMs ($\mathsf{ORam}_i$ for $i \ge 1$), we use a \textit{small block} size of $\chi \log_2 N$ bits for some constant $\chi \geq 2$.

The client storage is $O(\log^2 N) \cdot \omega(1)$ small blocks and $O(\log N) \cdot \omega(1)$ regular blocks.  Hence,
the total client storage is $O(\log N) \cdot \omega(1)$ regular blocks. Each operation
involves communication of $O(\log^2 N)$ small blocks and
$O(\log N)$ regular blocks.  Hence, the bandwidth cost (with respect to a regular block)
is $O(\log N)$.

\vspace{4mm}

As can be seen, using non-uniform block sizes leads to a much more efficient parametrization of recursive Path ORAM in terms of bandwidth. However, using uniform block sizes reduces the number of recursion levels, leading to a smaller round-trip response time.

\subsection{Shared Stash}
\label{sec:common_stash}

In Section~\ref{sec:SharedStashBounds}, we show that the client storage for our Recursive Path ORAM construction (with non-uniform block sizes) can be reduced from $O(\log^2 N)\cdot\omega(1)$ to $O(\log N)\cdot\omega(1)$ by having a \textit{single stash} shared among all levels of the recursion. This is possible while still maintaining a negligible probability of stash overflow.

\newcommand{\oflow}{R}
\newcommand{\oram}{\ensuremath{\mathsf{ORAM}}}
\newcommand{\st}{\ensuremath{\mathsf{st}}}
\newcommand{\ut}{\ensuremath{\mathsf{u}}}

\section{Bounds on Stash Usage}\label{sec:fail}

In this section we will analyze the stash usage for a non-recursive Path-ORAM, 
where each bucket in the Path-ORAM binary tree
stores a \emph{constant} number of blocks.
In particular, 
we analyze the probability that,
after a sequence of load/store operations, the number of blocks in the
the stash exceeds $R$, and show that
this probability decreases exponentially in $R$.

By $\oram_L^Z$ we denote a non-recursive Path-ORAM with $L+1$ levels in which each bucket stores $Z$ real/dummy blocks; the root is at level $0$ and the leaves are at level $L$. 

We define a {\em sequence of load/store operations} $\mbox{{\bf s}}$ as a triple $(\mbox{{\bf a}},\mbox{{\bf x}}, \mbox{{\bf y}})$ that contains (1) the sequence $\mbox{{\bf a}}=(a_i)_{i=1}^s$ of block addresses of blocks that are loaded/stored, (2) the sequence of labels $\mbox{{\bf x}}=(x_i)_{i=1}^s$ as seen by the server (line~\ref{line:GetPosition} in Figure~\ref{fig:Access}), and (3) the sequence $\mbox{{\bf y}}=(y_i)_{i= 1}^s$ of remapped leaf labels (line~\ref{line:SetPosition} in Figure~\ref{fig:Access}). The tuple $(a_1,x_1,y_1)$ corresponds to the most recent load/store operation, $(a_2,x_2,y_2)$ corresponds to the next most recent load/store operation, and so on.
A path from the root to some $x_i$ is known as an \emph{eviction} path,
and a path from the root to some $y_i$ is known as an \emph{assigned} path.
The number of load/store operations is denoted by $s=|\mbox{{\bf s}}|$. The {\em working set} corresponding to $\mbox{{\bf a}}$ is defined as the number of distinct block addresses $a_i$ in $\mbox{{\bf a}}$. We write $a(\mbox{{\bf s}})=\mbox{{\bf a}}$.

By $\oram_L^Z[\mbox{{\bf s}}]$ we denote the distribution of real blocks in $\oram_L^Z$ after a sequence $\mbox{{\bf s}}$ of load/store operations starting with an empty ORAM; the sequence $\mbox{{\bf s}}$ completely defines all the randomness needed to determine, for each block address $a$, its leaf label and which bucket/stash stores the block that corresponds to $a$. In particular the {\em number} of real blocks  stored in the buckets/stash can be reconstructed. 

We assume an infinite stash and in our analysis we investigate the usage $\st(\oram_L^Z[\mbox{{\bf s}}])$ of the stash defined as the number of real blocks that are stored in the stash after a sequence $\mbox{{\bf s}}$ of load/store operations.  In practice the stash is limited to some size $R$ and Path-ORAM fails after a sequence $\mbox{{\bf s}}$ of load/store operations if the stash needs more space: this happens if and only if the usage of the infinite stash is at least $\st(\oram_L^Z[\mbox{{\bf s}}])>R$.

\begin{theorem}[Main] \label{thm:main}
Let $\mbox{{\bf a}}$ be any sequence of block addresses with a working set of size at most $N$.
For a bucket size $Z=5$, tree height $L = \lceil \log N \rceil$ and stash size $R$, the probability of a Path ORAM failure after a sequence of load/store operations corresponding to $\mbox{{\bf a}}$, is at most 
\begin{eqnarray*}
&& \Pr(\st(\oram_L^5[\mbox{{\bf s}}])>R \,| \,a(\mbox{{\bf s}})=\mbox{{\bf a}}) \leq 14\cdot (0.6002)^R,
\end{eqnarray*}
where the probability is over the randomness that determines $\mbox{{\bf x}}$ and $\mbox{{\bf y}}$ in $\mbox{{\bf s}}=(\mbox{{\bf a}},\mbox{{\bf x}}, \mbox{{\bf y}})$.
\end{theorem}

As a corollary,  for $s$ load/store operations on $N$ data blocks, Path ORAM with client storage $\leq R$ blocks, server storage $20N$ blocks and bandwidth $10\log N$ blocks per load/store operation, fails during one of the $s$ load/store operations with probability $\leq s\cdot 14\cdot 0.6002^R$. 
So, if we assume the number of load/stores is equal to $s=poly(N)$, then, 
for a stash of size $O(\log N)\omega(1)$, the probability of Path ORAM failure during one of the load/store operations is negligible in $N$.

\vspace{.2cm}

\noindent
{\bf Proof outline.}
The proof of the main theorem consists of several steps: First, we introduce a second ORAM, called $\infty$-ORAM, together with an algorithm that post-processes the stash and buckets of $\infty$-ORAM in such a way that if $\infty$-ORAM gets accessed by a sequence $\mbox{{\bf s}}$ of load/store operations, then the process leads to a distribution of real blocks over buckets that is exactly the same as the distribution as in Path ORAM after being accessed by $\mbox{{\bf s}}$. 

Second, we characterize the distributions of real blocks over buckets in a $\infty$-ORAM for which post-processing leads to a stash usage $>R$. We show that the stash usage after post-processing is $>R$ if and only if there exists a subtree $T$ for which its ``usage'' in $\infty$-ORAM is more than its ``capacity''. This means that we can use the union bound to upper bound $\Pr[\st(\oram_L^Z[\mbox{{\bf s}}])>R \,| \,a[\mbox{{\bf s}}]=\mbox{{\bf a }})$ as a sum of probabilities over subtrees.

Third, we analyze the usage of subtrees $T$. We show how a mixture of a binomial and a geometric probability distribution expresses the probability of the number of real blocks that do not get evicted from $T$ after a sequence $\mbox{{\bf s}}$ of load/store operations. By using measure concentration techniques we prove the main theorem.

\vspace{-1mm}
\subsection{$\infty$-ORAM} \label{infty-oram}

We define $\infty$-ORAM, denoted by $\oram_L^\infty$, as an ORAM that exhibits the same tree structure as Path-ORAM with $L+1$ levels but where each bucket has an {\em infinite} size.
The $\infty$-ORAM is used only as a tool for usage analysis, and does not
need to respect any security notions.

In order to use $\oram_L^\infty$ to analyze the stash usage of $\oram_L^Z$,
we define a {\em post-processing} greedy algorithm
$G_Z$ that takes
as input the state of $\oram_L^\infty$ after a sequence $\mbox{{\bf s}}$ of load/store operations and attempts to reassign blocks such that each bucket stores at most $Z$ blocks, putting
excess blocks in the (initially empty) stash if necessary.
We use $\st^Z(\oram_L^\infty[\mbox{{\bf s}}])$ to denote the stash usage after the greedy algorithm is applied.
The algorithm repeats the following steps
until there is no bucket that stores more than $Z$ blocks, where the stash is initially empty.

\begin{enumerate}
\item Select a block in a bucket that stores more than $Z$ blocks. Suppose 
the bucket is at level $h$, and $P$ is the path from the bucket to the root.

\item Find the highest level $i\leq h$ such that the bucket at level $i$ on the path $P$ stores less than $Z$ blocks. 
If such a bucket exists, then use it to store the block. If it does not exist, then put the block in the stash. 
\end{enumerate}

\begin{lemma} \label{lem1} The stash usage in a post-processed $\infty$-ORAM is exactly the same as the stash usage in Path-ORAM:

$$ \st^Z(\oram_L^\infty[\mbox{{\bf s}}]) = \st(\oram_L^Z[\mbox{{\bf s}}]).$$
\end{lemma}

\noindent
\begin{proof} We first notice that the order
 in which the greedy strategy $G_Z$ processes blocks from the stash 
 does not affect the number of real blocks in each bucket that are stored  
 in server storage after post-processing.
 
Suppose the greedy strategy first processes a block $b_1$ with label $x_1$ 
stored in a bucket at level $h_1$  and suppose it finds empty space in server storage at level $i_1$; 
the greedy strategy up to block $b_1$ has used up all the empty space in server storage 
allocated to the buckets along the path to the leaf with label $x_1$ at levels $i_1 < i \leq h_1$. 
Suppose next a stash block $b_2$ with label $x_2$ stored in a bucket at level $h_2$ finds empty space in server storage at level $i_2$; the greedy strategy up to block $b_2$, which includes the post-processing of $b_1$, has used up all the empty space in server storage allocated to the buckets along the path to the leaf with label $x_2$ at levels $i_2<i\leq h_2$.  
If we swap the post-processing of $b_1$ and $b_2$, then $b_2$ is able to find empty space in the bucket at level $i_2$, but may find empty space at a higher level since $b_1$ has not yet been processed. In this case, $i_2<i_1$ and paths $x_1$ and $x_2$ intersect at least up to and including level $i_1$. This means that $b_2$ is stored at level $i_1$ and $b_1$ will use the empty space in server storage at level $i_2$. This means that the number of blocks that are stored in the different buckets after post-processing $b_1$ and $b_2$ is the same if $b_1$ is processed before or after $b_2$.

Now, we are able to show, by using induction in $|\mbox{{\bf s}}|$, that
for each bucket $\mathsf{b}$ in a post-processed $\infty$-ORAM after a sequence 
$\mbox{{\bf s}}$ of load/store operations, the number of real blocks stored in $\mathsf{b}$ that are in server storage is equal to the number of blocks stored in the equivalent bucket in Path-ORAM after applying the same sequence $\mbox{{\bf s}}$ of load/store operations: The statement clearly holds for $|\mbox{{\bf s}}|=0$ when both ORAMs are empty. Suppose it holds for $|\mbox{{\bf s}}|\geq 0$. Consider the next load/store operation to some leaf $f$. After reading the path to leaf $f$ into the cache, Path-ORAM moves blocks from its cache/stash into the path from root to leaf $f$ according to algorithm $G_Z$. Therefore, post-processing after the first $|\mbox{{\bf s}}|$ operations followed by the greedy approach of Path ORAM that processes the  $(|\mbox{{\bf s}}|+1)$-th operation is equivalent to post-processing after $|\mbox{{\bf s}}|+1$ operations where some blocks may be post-processed twice. Since the order in which blocks are post-processed does not matter, we may group together the multiple times a block $b$ is being post-processed and this is equivalent to post-processing $b$ exactly once as in $G_Z$. The number of real blocks in the stash is the same and this proves the lemma.
\hfill \end{proof}

\subsection{Usage/Capacity Bounds} \label{failure}

To investigate when
a not processed $\oram_L^\infty$ can lead to a stash usage of $>R$ after post-processing, we start by analyzing bucket usage over subtrees.
When we talk about a subtree $T$ of the binary tree, we always implicitly assume that it contains the root of the ORAM tree; in particular, if a node is contained in $T$, then so are all its ancestors.
We define $n(T)$ to be the total number of nodes in $T$.
For $\infty$-ORAM we define the usage $\ut^T(\oram_L^\infty[\mbox{{\bf s}}])$ of $T$ after a sequence $\mbox{{\bf s}}$ of load/store operations as the actual number of real blocks that are stored in the buckets of $T$.

The following lemma characterizes the stash usage:

\begin{lemma} \label{lem2}
The stash usage $\st^Z(\oram_L^\infty[\mbox{{\bf s}}])$ in post-processed $\infty$-ORAM is $> R$ if and only if there exists a subtree $T$ in $\oram_L^\infty$ such that $\ut^T(\oram_L^\infty[\mbox{{\bf s}}])> n(T)\cdot Z+R.$
\end{lemma}

\noindent
\begin{proof} 

\noindent \emph{If part:} Suppose $T$ is a subtree such that 
$\ut^T(\oram_L^\infty[\mbox{{\bf s}}])> n(T)\cdot Z+R$. Observe that
the greedy algorithm can assign
the blocks in a bucket only to an ancestor bucket.
Since $T$ can store at most $n(T) \cdot Z$ blocks, more than $R$ blocks
must be assigned to the stash by the greedy algorithm $G_Z$.

\noindent \emph{Only if part:} Suppose that $\st^Z(\oram_L^\infty[\mbox{{\bf s}}])>R$.   Define $T$
to be the maximal subtree that contains all buckets with exactly $Z$ blocks after post-processing
by the greedy algorithm $G_Z$.  Suppose $b$ is a bucket not in $T$.  By the maximality of $T$,
there is an ancestor (not necessarily proper ancestor) bucket $b'$ of $b$ that 
contains less than $Z$ blocks after post-processing,
which implies that no block from $b$ can go to the stash.
Hence, all blocks that are in the stash must have originated from a bucket in $T$.
Therefore, it follows that $\ut^T(\oram_L^\infty[\mbox{{\bf s}}])> n(T)\cdot Z+R$
\hfill \end{proof}

\begin{lemma}[Worst-case Address Pattern]
\label{lemma:worst_address}
Out of all address sequences ${\bf a}$ on a working set
of size $N$, the probability
$\Pr[\st^Z(\oram_L^\infty[\mbox{{\bf s}}])> R|a(\mbox{{\bf s}}) = {\bf a}]$
is maximized by a sequence ${\bf a}$
in which each block address appears exactly once, i.e.,
$s=N$ and there are no duplicate block addresses.
As a consequence, for such an address pattern, the labels in $(x_i)_{i=1}^N$ and $(y_i)_{i=1}^N$ are all statistically independent of one another.
\end{lemma}

\begin{proof} 
Suppose that there exists an address in $\mbox{{\bf a}}$ that has been loaded/stored twice in $\infty$-ORAM.
Then, there exist indices $i$ and $j$, $i<j$, with $a_i=a_j$. Without the $j$-th load/store, the working set remains the same  and it is more likely for older blocks corresponding to $a_k$, $k>j$ to {\em not} have been evicted from $T$ (since there is one less load/store that could have evicted an older block to a higher level outside $T$; also notice that buckets in $\infty$-ORAM are infinitely sized, so, removing the $j$-th load/store does not generate extra space that can be used for storage of older blocks that otherwise would not have found space). So, the probability
$\Pr[\st^Z(\oram_L^\infty[\mbox{{\bf s}}])> R|a(\mbox{{\bf s}}) = {\bf a}]$
is maximized when
$\mbox{{\bf a}}=(a_i)_{i=1}^s$ is a sequence of block addresses without duplicates.
\hfill \end{proof}

\noindent \textbf{Bounding Usage for Each Subtree.}
In view of Lemma~\ref{lemma:worst_address},
we fix a sequence
$\mbox{{\bf a}}$ of $N$ distinct block addresses.
The randomness comes from the independent choices of labels
$(x_i)_{i=1}^N$ and $(y_i)_{i=1}^N$.

As a corollary to Lemmas \ref{lem1} and \ref{lem2}, we obtain
\begin{eqnarray*}
&& \Pr[\st(\oram_L^Z[\mbox{{\bf s}}])>R] \\
&=& \Pr[\st^Z(\oram_L^\infty[\mbox{{\bf s}}])> R] \\
&=& \Pr[ \exists T \ \ut^T(\oram_L^\infty[\mbox{{\bf s}}])> n(T)Z+R ]\\
&\leq& \sum_T \Pr[\ut^T(\oram_L^\infty[\mbox{{\bf s}}])> n(T)Z+R],
\end{eqnarray*}
where $T$ ranges over all subtrees containing the root, and the inequality follows from the union bound.

Since the number of ordered binary trees of size $n$ is equal to the Catalan number $C_n$, which is $\leq 4^n$,

\begin{eqnarray*} \label{eq:union}
& \Pr[\st(\oram_L^Z[\mbox{{\bf s}}])>R] \\
& \leq  \sum_{n\geq 1} 4^n \max_{T:n(T)=n} \Pr[\ut^T(\oram_L^\infty[\mbox{{\bf s}}])> nZ+R].
\end{eqnarray*}

We next give a uniform upper bound for
$$\Pr[\ut^T(\oram_L^\infty[\mbox{{\bf s}}])> nZ+R]$$
in terms of $n$, $Z$ and $R$.

\subsection{Usage of Subtree via Eviction Game}

In view of Lemma~\ref{lemma:worst_address},
we assume that the request sequence is of length $N$,
and consists of distinct block addresses.
Recall that the $2N$ labels in $\vector{x}= (x_i)_{i=1}^N$
and $\vector{y}= (y_i)_{i=1}^N$ are independent,
where each label corresponds to a leaf bucket.
The indices $i$ are given in reversed order, i.e, $i=N$ is the first access
and $i=1$ is the last access.

At every time step $i$, a block is requested. The block resides
in a random path with some label $x_i$ previously chosen (however the random choice has never been revealed).
This path is now visited. 
Henceforth, this path is called the {\it eviction path} $P_{\text{evict}}(i)$.

Then, the block is logically assigned to  
a freshly chosen random path with label $y_i$, referred to as the {\it assigned path} $P_{\text{assign}}(i)$.

Recall that given subtree $T$, we want to probabilistically bound 
the quantity $u^T(\oram_L^\infty[\vector{s}])$, which is the number
of blocks that, at the end of an access sequence $\vector{s}$,
survive in the subtree $T$, i.e., have not been evicted out of the subtree $T$.

\begin{definition}[Exit node]
For a given path $P$ leading to some leaf node, 
suppose that some node $u$ is the first node of the path $P$ 
that is not part of $T$, then we refer to node $u$ as the exit node,
denoted $u := {\sf exit}(P, T)$. If the whole path $P$ is contained
in $T$, then the exit node ${\sf exit}(P, T)$ is \emph{null}.
\end{definition}

To bound the number of surviving blocks in $T$ at the end of 
an access sequence of $N$,
we focus on an individual block that is requested at step
$i$ in this sequence. 
Suppose this block 
resides in path $P_{\text{evict}}(i)$, and
is assigned to some path $P_{\text{assign}}(i)$ at the end of the request.
Observe that this block will be evicted from tree $T$ at the end of the sequence
\emph{iff} the following holds:
there exists a later step $j \leq i$ (recalling that smaller indices mean later) such that
both exit nodes ${\sf exit}(P_{\text{evict}}(j), T)$ and  ${\sf exit}(P_{\text{assign}}(i), T)$
are equal and not null.
Otherwise the block will survive in tree $T$ at the end of the sequence ---
and will contribute  
to the count of the total surviving blocks.

We next define some notations relating to the subtree $T$.

Let $F$ be the set of nodes in $T$ that are also leaves of the
ORAM binary tree; we denote $l := |F|$.
We augment the tree $T$ by adding nodes to form $\widehat{T}$ in the following way.
If a node in $T$ has any child node $v$ that is not in $T$,
then node $v$ will be added to $\widehat{T}$.
The added nodes 
in $\widehat{T}$ are referred to as exit nodes,
denoted by $E$; the leaves of $\widehat{T}$ are denoted by
$\widehat{E} = E \cup F$.  Observe that if $T$ contains $n$ nodes
and $|F|=l$, then $|\widehat{E}| = n - l + 1$.

We summarize the notations we use in Table~\ref{tab:ProofNotations}.

\vspace{3pt}

\begin{table*}[t]
\centering
\begin{tabular}{l|l}
Variable & Meaning\\
\hline
$T$ & a subtree rooted at the root of the ${\sf ORAM}_\infty$
binary tree\\
$\widehat{T}$ & augmented tree by including every child (if any) of every node in $T$  \\
$F$ & nodes of a subtree $T$ that are leaves to the ORAM binary tree\\
$E$ & set of exit nodes of a subtree $T$\\
$\widehat{E} := E \cup F$ &set of all leaves of $\widehat{T}$\\
$Z$ & capacity of each bucket\\
\end{tabular}
\caption{{\bf Notations for the analysis.}}
\label{tab:ProofNotations}
\end{table*}

\begin{figure}
\centering
\includegraphics[width=\columnwidth]{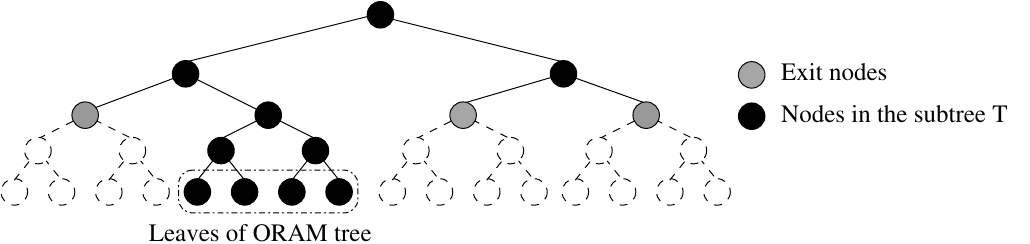}
\caption{A subtree containing some leaves of the original
ORAM binary tree, augmented with the exit nodes}
\end{figure}

\noindent \textbf{Eviction Game.} Consider the following eviction game, which is essentially 
playing the ORAM access sequence
{\it reversed in time}.  Initially, all nodes in $\widehat{E}$
are marked \emph{closed}; all nodes in $F$ will remain closed, and only nodes in $E$ may
be marked \emph{open} later.

For $i = 1,  2, \ldots, N$, do the following:
\begin{enumerate}
\item Pick a random \emph{eviction path} starting from the root, denoted
$P_{\text{evict}}(i)$.  The path
will intersect exactly one node $v$ in $\widehat{E}$.
If $v$ is an exit node in $E$, mark node $v$ as open (if
that exit node is already marked open, then it will continue to be open).

\item For a block requested in time step $i$, 
pick a random assignment path $P_{\text{assign}}(i)$ starting from
the root.  The path $P_{\text{assign}}(i)$ 
will intersect exactly one node in $\widehat{E}$.
If this node is closed, then the block \emph{survives}; otherwise, the block is 
\emph{evicted} from $T$.
\end{enumerate}

For $i \in [N]$, a block $b_i$ from step $i$ survives because its corresponding assignment path intersects
a node in $F$, or an exit node that is closed.
We define $X_i$ to be the indicator
variable that equals 1 \emph{iff} the path $P_{\text{assign}}(i)$
intersects a node in $F$ (and block $b_i$ survives), $Y_i$ to be the indicator variable
that equals 1 \emph{iff} the path $P_{\text{assign}}(i)$
intersects a node in $E$ and block $b_i$ survives.
Define $X := \sum_{i \in [N]} X_i$ and $Y := \sum_{i \in [N]} Y_i$;
observe that the number of blocks that survive is $X + Y$.

\subsection{Negative Association}

Independence is often required to show measure concentration.  However,
the random variables $X$ and $Y$ are not independent.
Fortunately, $X$ and $Y$ are negatively associated.  For simplicity,
we show a special case in the following lemma, which will be useful in the later
measure concentration argument.

\begin{lemma}
For $X$ and $Y$ defined above, and all $t \geq 0$, $E[e^{t(X+Y)}] \leq
E[e^{tX}] \cdot E[e^{tY}]$.
\end{lemma}

\begin{proof}
Observe that there exists some $p \in [0,1]$, such that
for all $i \in [N]$, $p = \Pr[X_i = 1]$.  For each $i \in [N]$,
define $q_i := \Pr[Y_i = 1]$.  Observe that the $q_i$'s are
random variables, and they are non-increasing in $i$ and determined
by the choice of eviction paths.

For each $i \in [N]$, $X_i + Y_i \leq 1$, and hence
$X_i$ and $Y_i$ are negatively associated.
Conditioning on  $\vector{q} := (q_i)_{i \in [N]}$, observe that the
$(X_i, Y_i)$'s are determined by the choice of independent
assignment paths in different rounds, and hence are
independent over different $i$'s.
Hence, it follows from \cite[Proposition 7(1)]{DubhashiR98} that conditioning on $\vector{q}$, $X$ and $Y$ 
are negatively associated.

In particular, for non-negative $t$, we have $E[e^{t(X+Y)}|\vector{q}] \leq E[e^{tX}|\vector{q}] \cdot E[e^{tY}|\vector{q}] = E[e^{tX}] \cdot E[e^{tY}|\vector{q}]$,
where the last equality follows because $X$ is independent of $\vector{q}$.
Taking expectation over $\vector{q}$ gives $E[e^{t(X+Y)}] \leq
E[e^{tX}] \cdot E[e^{tY}]$. 
\end{proof}

\subsection{Stochastic Dominance}

Because of negative association, we consider 
two games to analyze the random variables $X$ and $Y$ separately.
The first game is a balls-and-bins game which 
produces a random variable $\widehat{X}$ that
has the same distribution as $X$.
The second game is a modified eviction game 
with countably infinite number of rounds, which produces
a random variable $\widehat{Y}$ that stochastically dominates
$Y$, in the sense that
$\widehat{Y}$ and $Y$ can be coupled such that $Y \leq \widehat{Y}$.

\begin{enumerate}
\item
{\bf Balls-and-Bins Game:} For simplicity, we assume $N = 2^L$.
In this game, $N$ blocks
are thrown independently and uniformly at random into $N$ buckets 
corresponding to the leaves of the
ORAM binary tree.  
The blocks that fall in the leaves in $F$ survive.
Observe that the number $\widehat{X}$ of surviving blocks
has the same distribution as $X$.

\item
{\bf Infinite Eviction Game:}
This is the same as before, except for the following differences.

\begin{enumerate}
\item Every node in $\widehat{E}$ is initially closed, but
all of them could be open later.  In particular, if
some eviction path $P_{\text{evict}}(i)$ intersects a leaf node
$v \in F$, node $v$ will become open. 

\item There could be countably infinite number of rounds, until eventually all
nodes in $\widehat{E}$ are open, and no more blocks can survive after that.
\end{enumerate}

Let $\widehat{Y}$ be the total number of surviving blocks
in the infinite eviction game.
Observe that a block with assigned path intersecting a node in $F$
will not be counted towards $Y$, but might be counted towards $\widehat{Y}$
if the node is closed.
Hence, there is a natural coupling such that
the number of surviving blocks in the first $N$ rounds
in the infinite game
is at least $Y$ in the finite eviction game.  Hence,
the random variable $\widehat{Y}$ stochastically dominates ${Y}$.
Hence, we have for all non-negative $t$, $E[e^{t Y}] 
\leq E[e^{t \widehat{Y}}]$.

\end{enumerate}

\subsection{Measure Concentration for the Number of Surviving Blocks in Subtree}

We shall find parameters $Z$ and $\oflow$ such that,
with high probability, $X + Y$ is at most $n Z + \oflow$.
For simplicity, we assume $N = 2^L$ is a power of two.

\begin{lemma} \label{lemma:meas_conc}
Suppose that the address sequence is of length $N$, where $L := \ceil{\log_2 N}$.
Moreover, suppose that $T$ is a subtree of the binary ORAM tree containing the root having $n = n(T)$ nodes.
Then, for $Z=5$, for any $\oflow > 0$, 
$\Pr[\ut^T(\oram_L^\infty[\vector{s}]) > n \cdot Z + R] \leq
\frac{1}{4^n} \cdot (0.9332)^n \cdot e^{-0.5105 \oflow}$.
\end{lemma}

\begin{proof}
Because of negative association between $X$ and $Y$, and stochastic
dominance by $\widehat{X}$ and $\widehat{Y}$, we analyze $\widehat{X}$ and
$\widehat{Y}$.

Observe that if $T$ has $n$ nodes, among which $l$ are in $F$,
then $\widehat{T}$ has $n-l+1$ leaves in $\widehat{E}$.

\vspace{5pt}

\noindent \textbf{Balls-and-Bins Game.}
Recall that $X_i$ is the indicator variable for whether
the assigned path $P_{\text{assign}}(i)$
intersects a bucket in $F$.
Then, $\widehat{X} = \sum_{i \in [n]} X_i$.
Recall that $l = |F|$ and there are $N = 2^L$ leaf buckets in the binary ORAM tree.
Observe that for real $t$,
$E[e^{t X_i}] = (1 - \frac{l}{N}) + \frac{l}{N} e^t  \leq \exp(\frac{l}{N}(e^t-1))$,
and hence by independence,
$E[e^{t\widehat{X}}] \leq \exp(l (e^t -1))$.

\vspace{5pt}

\noindent \textbf{Infinite Eviction Game.}
For each $v \in \widehat{E}$, 
suppose $v$ is at depth $d_v$ (the root is at depth 0), and let its weight be $w(v) := \frac{1}{2^{d_v}}$.  Observe
that the sum of weights of nodes in $\widehat{E}$ is 1.

Define $M_j$ to be the number of surviving blocks
such that 
there are exactly $j$ open nodes at the moment when the corresponding assigned paths are chosen.
Since $\widehat{E}$ contains $n-l+1$ nodes, we consider $j$ from 1 to $k=n-l$.
(Observe that for in the first round of the infinite eviction game, the first eviction path closes one of the nodes
in $\widehat{E}$, and hence, the number of remaining open nodes for the first block is $n-l$.)
Let $Q_j$ be the number of rounds in which, at the moment
when the assigned path is chosen, there are exactly $j$ open nodes.
Let $w_j$ be the weight of the $j$th open node.
Define $q_j := 1 - \sum_{i=1}^j w_i$.  Observe that conditioning on $q_j$,
$Q_j$ follows the geometric distribution with parameter $q_j$;
moreover, conditioning on $q_j$ and $Q_j$, $M_j$ is the sum
of $Q_j$ independent Bernoulli random variables, each with expectation
$q_j$.

Define $\widehat{Y} := \sum_{j=1}^k M_j$.  We shall analyze the moment generating
function $E[e^{t \widehat{Y}}]$ for appropriate values of $t > 0$.  The strategy is
that we first derive an upper bound $\phi(t)$ for $E[e^{t M_k} | \vector{w}]$,
where $\vector{w} := \{w_j\}_{j \leq k}$,
that depends only on $t$ (and in particular independent of $\vector{w}$ or $k$).
This allows us to conclude that $E[e^{t\widehat{Y}}] \leq (\phi(t))^k$.

For simplicity, in the below argument, we assume that
we have conditioned on $\vector{w} = \{w_j\}_{j \leq k}$ and we write
$Q = Q_k$, $M = M_k$ and $q = q_k$.

Recall that $M$ is a sum of $Q$ independent Bernoulli random variables,
where $Q$ has a geometric distribution.  Therefore, we have the following.

\begin{eqnarray}
E[e^{t M} | \vector{w}] & = & \sum_{i \geq 1} q(1-q)^i E[e^{t M} | Q = i, \vector{w}] \label{eq:0} \\ 
& \leq & \sum_{i \geq 1} q(1-q)^{i-1} \exp(q i(e^t-1)) \label{eq:0b} \\
& = & \frac{q \exp(q(e^t-1))}{1 - (1-q) \exp(q(e^t-1))} \label{eqn:1} \\
& = & \frac{q }{\exp(-q(e^t-1)) - (1-q)} \\
& \leq & \frac{q}{1 - q(e^t-1) -1 + q} \label{eqn:2} \\
& = & \frac{1}{2 - e^t}.
\end{eqnarray}

From (\ref{eq:0}) to (\ref{eq:0b}),
we consider the moment generating function
of a sum of $i$ independent Bernoulli random variables,
each having expectation $q$:
$E[e^{t M} | Q = i, \vector{w}] = ((1 - q) + q e^t)^i \leq \exp(q i(e^t-1))$.
In (\ref{eqn:1}), for the series to converge,
we observe that $(1-q) \exp(q(e^t-1)) \leq \exp(q(e^t-2))$,
which is smaller than 1 when $0 < t < \ln 2$.
In (\ref{eqn:2}), we use $1 - u \leq \exp(-u)$ for all real $u$.

Hence, we have shown that for $0 < t < \ln 2$,
$E[e^{t \sum_{j=1}^k M_j} | \vector{w}] = 
E[e^{t \sum_{j=1}^{k-1} M_j} | \vector{w}]
\cdot E[e^{t M_k} | \vector{w}]$ (since conditioning on $\vector{w}$,
$M_k$ is independent of past history),
which we show from above is at most
$E[e^{t \sum_{j=1}^{k-1} M_j} | \vector{w}] \cdot
(\frac{1}{2 - e^t})$.  Taking expectation over $\vector{w}$ gives
$E[e^{t \sum_{j=1}^k M_j}]
\leq E[e^{t \sum_{j=1}^{k-1} M_j}] \cdot
(\frac{1}{2 - e^t})$.

Observe that the inequality holds independent of the value
of $q = q_k$.  Therefore, the same argument can be used to prove that,
for any $1 < \kappa \leq k$,
we have $E[e^{t \sum_{j=1}^\kappa M_j}]
\leq E[e^{t \sum_{j=1}^{\kappa-1} M_j}] \cdot
(\frac{1}{2 - e^t})$.

Hence, a simple induction argument on $k=n-l$ can show that
$E[e^{t \widehat{Y}}] \leq (\frac{1}{2 - e^t})^{n-l}$.

\vspace{5pt}

\noindent \textbf{Combining Together.}
Let $Z$ be the capacity for each bucket in
the binary ORAM tree, and
$\oflow$ be the number blocks overflowing from the binary tree
that need to be stored in the stash.

We next perform a standard moment generating function argument.
For $0 < t < \ln 2$,
$\Pr[X + Y \geq n Z  + \oflow]
= \Pr[e^{t(X+Y)} \geq e^{t(n Z  + \oflow)}]$,
which by Markov's Inequality, is at most
$E[e^{t(X+Y)}] \cdot e^{-t(n Z  + \oflow)}$,
which, by negative associativity and stochastic dominance,
is at most 
$E[e^{t \widehat{X}}] \cdot
E[e^{t \widehat{Y}}] \cdot e^{-t(n Z  + \oflow)}
\leq  (\frac{e^{-t Z}}{2 - e^t})^{n-l} \cdot
e^{l(e^t - 1 - t Z)} \cdot e^{-t \oflow}$.

Putting $Z = 5$, and $t = 0.5105$,
one can check that $\max \{\frac{e^{-t Z}}{2 - e^t}, e^{e^t - 1 - t Z}\} \leq \frac{1}{4} \cdot 0.9332$, and so
the probability is
at most $\frac{1}{4^n} \cdot (0.9332)^n \cdot e^{-0.5105 \oflow}$.
\end{proof}

\noindent
{\bf Proof of Theorem~\ref{thm:main}.}
By applying Lemma~\ref{lemma:meas_conc} to inequality (\ref{eq:union}),
we have the following:
$ \Pr[\st(\oram_L^Z[\mbox{{\bf s}}])>R|a(\mbox{{\bf s}})=\mbox{{\bf a}}] 
\leq \sum_{n\geq 1} 4^n \cdot \frac{1}{4^n} \cdot (0.9332)^n \cdot e^{-0.5105 \oflow}
\leq  14 \cdot (0.6002)^\oflow$,
as required.

\vspace{10pt}

\subsection{Bounds for Shared Stash}
\label{sec:SharedStashBounds}

We now show that if all levels of the recursion use the same stash, the stash size is $O(\log N)\cdot\omega(1)$ with high probability.

Suppose there are $K = O(\log N)$ levels of recursion in the recursive Path ORAM.
We consider a moment after a sequence of ORAM operations are executed. 
For $k \in [K]$, let $S_k$ be the number of blocks in the stash from level $k$.
From Theorem~\ref{thm:main}, for each $k \in [K]$, for each $R > 0$, 
$\Pr[S_k > R] \leq 14 \cdot (0.6002)^R$. Observing that a geometric distribution
$G$ with parameter $p$ satisfies $\Pr[G > R] \leq (1 - p)^R$, we
have the following.

\begin{proposition} \label{prop:geom_dom}
For each $k \in [K]$, the random variable $S_k$ is stochastically dominated by
$3 + G$, where $G$ is the geometric distribution with parameter $p = 1 - 0.6002 = 0.3998$.
\end{proposition}

From Proposition~\ref{prop:geom_dom}, it follows
that the number of stash blocks in the common storage is stochastically dominated
by $3K + \sum_{k \in [K]} G_k$, where $G_k$'s are independent geometric distribution
with parameter $p = 0.3998$.  It suffices to perform a standard measure concentration
analysis on a sum of independent geometrically distributed random variables, but 
we need to consider the case that the sum deviates significantly from its mean, 
because we want to achieve negligible failure probability.

\begin{lemma}[Sum of Independent Geometric Distributions]
\label{lemma:sum_geom}
Suppose $(G_k: k \in [K])$ are independent geometrically distributed random variables
with parameter $p \in (0,1)$.  Then, for $R > 0$, we have
$\Pr[\sum_{k\in[K]} G_k >  E[\sum_{k\in[K]} G_k] + R] \leq \exp(-\frac{p R}{2} + \frac{K}{2})$.
\end{lemma}

\begin{proof}
We use the standard method of moment generating function.  For $t \leq \frac{p}{2}$,
we have, for each $k \in [K]$,

\begin{eqnarray}
E[e^{tG_k}] & = & \sum_{i \geq 1} p(1-p)^i \cdot e^{it} \\
& = & \frac{p e^t}{1 - (1-p)e^t} = \frac{p}{p + e^{-t} - 1}  \label{g:0} \\
& \leq & \frac{p}{p - t} = \frac{1}{1 - \frac{t}{p}} \label{g:1}  \\
& \leq & 1 + \frac{t}{p} + \frac{2 t^2}{p^2} \label{g:2}\\
& \leq & \exp(\frac{t}{p} + \frac{2 t^2}{p^2}),
\end{eqnarray}

where in (\ref{g:0}), the geometric series converges, because
$t \leq \frac{p}{2} < \ln \frac{1}{1-p}$, for $0 < p < 1$.
In (\ref{g:1}), we use the inequality $1 - e^{-t} \leq t$;
in (\ref{g:2}), we use the inequality
$\frac{1}{1-u} \leq 1 + u + 2u^2$, for $u \leq \frac{1}{2}$.

Observing that $E[G_k] = \frac{1}{p}$, we have for $0 < t \leq \frac{p}{2}$,

$\Pr[\sum_{k\in[K]} G_k >  E[\sum_{k\in[K]} G_k] + R] \leq
E[\exp(t \sum_{k\in[K]} G_k)] \cdot \exp(-t(\frac{K}{p} + R))
\leq \exp(- tR + \frac{2 t^2 K}{p^2})$.
Putting $t = \frac{p}{2}$,
we have $\Pr[\sum_{k\in[K]} G_k >  E[\sum_{k\in[K]} G_k] + R] \leq \exp(-\frac{p R}{2} + \frac{K}{2})$, as required.
\end{proof}

From Lemma~\ref{lemma:sum_geom},
observing that $K = O(\log N)$ and $p = 0.3998$,
to achieve failure probability $\frac{1}{N^{\omega(1)}}$,
it suffices to set the capacity of the common stash storage
to be $\frac{1}{p} \cdot (O(K) + (\log N) \cdot \omega(1) ) = \Theta(\log N) \cdot \omega(1)$
blocks.

\section{Applications and Extensions}

\subsection{Oblivious Binary Search Tree}
Based on a class of recursive, binary tree based ORAM constructions, 
Gentry \etal propose a novel method for performing an entire binary search
using a single ORAM lookup~\cite{gentryoram}.
Their method is immediately applicable to Path ORAM.
As a result, Path ORAM can be used to perform search on an oblivious binary search tree,
using $O(\log^2 N)$ bandwidth. Note that since a binary search requires navigating a path of $O(\log N)$ nodes, using existing generic ORAM techniques would lead to bandwidth cost of $O((\log N)^3/\log\log N)$.

\subsection{Stateless ORAM}
Oblivious RAM is often considered in a single-client model, but it is sometimes useful to have multiple clients accessing the same ORAM. In that case, in order to avoid complicated (and possibly expensive) oblivious state synchronization between the clients, Goodrich~\etal introduce the concept of \textit{stateless} ORAM~\cite{GMOT12} where the client state is small enough so that any client accessing the ORAM can download it before each data access and upload it afterwards. Then, the only thing clients need to store is the private key for the ORAM (which does not change as the data in the ORAM changes).

In our recursive Path ORAM construction, we can download and upload the client state before and after each access. Since the client state is only ${O(\log N) \cdot \omega(1)}$ and the bandwidth is $O(\log N)$ when $B = \Omega(\log^2 N)$, we can reduce the permanent client state 
to $O(1)$ and achieve a bandwidth of $O(\log N) \cdot \omega(1)$. Note that \textit{during} an access the client still needs about $O(\log N) \cdot \omega(1)$ \textit{transient} client storage to perform the \access operation, but after the \access operation completes, the client only needs to store the private key.

For smaller blocks when $B = \Omega(\log N)$, we can achieve $O(1)$ permanent client storage, $O(\log N) \cdot \omega(1)$ transient client storage, and $O(\log^2 N)$ bandwidth cost.

\subsection{Secure Processors}

In a secure processor setting, private computation is done inside a tamper-resistant processor (or board) and main memory (e.g., DRAM) accesses are vulnerable to eavesdropping and tampering. As mentioned earlier, Path ORAM is particularly amenable to hardware design because of its simplicity and low on-chip storage requirements.

Fletcher \etal~\cite{ascend-stc12,cwf-masters} and Ren \etal~\cite{REN13} built a simulator for a secure processor based on \name. They optimize the bandwidth cost and stash size of the recursive construction by using a smaller $Z=3$ in combination with a background eviction process that does not break the Path ORAM invariant.

Maas \etal~\cite{phantom} built a hardware implementation of a \name based secure processor using FPGAs and the Convey platform.

Ren \etal~\cite{REN13} and Maas \etal~\cite{phantom} report about 1.2X to 5X performance overhead for many benchmarks such as SPEC traces and SQLite queries. To achieve this high performance, these hardware Path ORAM designs rely on on-chip caches while making Path ORAM requests only when last-level cache misses occur.

\subsection{Integrity}
\label{sec:integrity}

Our protocol can be easily extended to provide integrity (with freshness) for every access to the untrusted server storage. 
Because data from untrusted storage is always fetched and stored in the form of a tree paths, we can achieve integrity by simply treating the Path ORAM tree as a Merkle tree where data is stored in all nodes of the tree (not just the leaf nodes). In other words, each node (bucket) of the Path ORAM tree is tagged with a hash of the following form
$$H(b_1 \parallel b_2 \parallel \ldots \parallel b_Z \parallel h_1 \parallel h_2)$$
where $b_i$ for $i \in \{1, 2, \ldots, Z\}$ are the blocks in the bucket (some of which could be dummy blocks) and $h_1$ and $h_2$ are the hashes of the left and right child. For leaf nodes, $h_1 = h_2 = 0$. Hence only two hashes (for the node and its sibling) need to be read or written for each \ReadBucket or \WriteBucket operation.

In \cite{oram-hpec13}, Ren \etal further optimize the integrity verification overhead for the recursive \name construction.

\section{Evaluation}
\label{sec:experiments}

In our experiments, the Path ORAM uses a binary tree with height $L = \ceil{\log_2(N)}-1$.

\subsection{Stash Occupancy Distribution} \label{sec:StashDistribution}

\begin{figure}[t]
\centering
\includegraphics[width=0.6\textwidth]{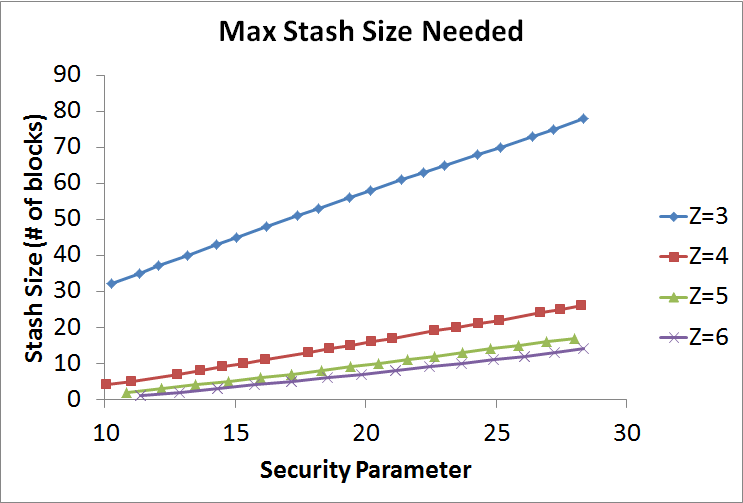}
\caption{Empirical estimation of the required stash size to achieve failure probability less than $2^{-\lambda}$ where $\lambda$ is the security parameter. Measured for $N=2^{16}$, but as Figure~\ref{fig:CacheSizes-VaryingN} shows, the stash size does not depend on $N$  (at least for $Z=4$).  The measurements represent a worst-case (in terms of stash size) access pattern. The stash size does not include the temporarily fetched path during \access.}
\label{fig:CacheSizes-LambdaVsStashSize}
\end{figure}

\paragraph{Stash occupancy}
{\it In both the experimental results and the theoretical analysis, 
we define the stash occupancy to be the  
number of overflowing blocks (i.e., the number of blocks that remain in the stash) after the write-back phase of each ORAM access. }
This represents the {\it persistent} 
local storage required on the client-side. 
In addition, the client also requires under $Z \log_2 N$ 
{\it transient} storage for temporarily caching a path fetched
from the server during each ORAM access.

\begin{figure}[t]
\centering
\includegraphics[width=0.6\textwidth]{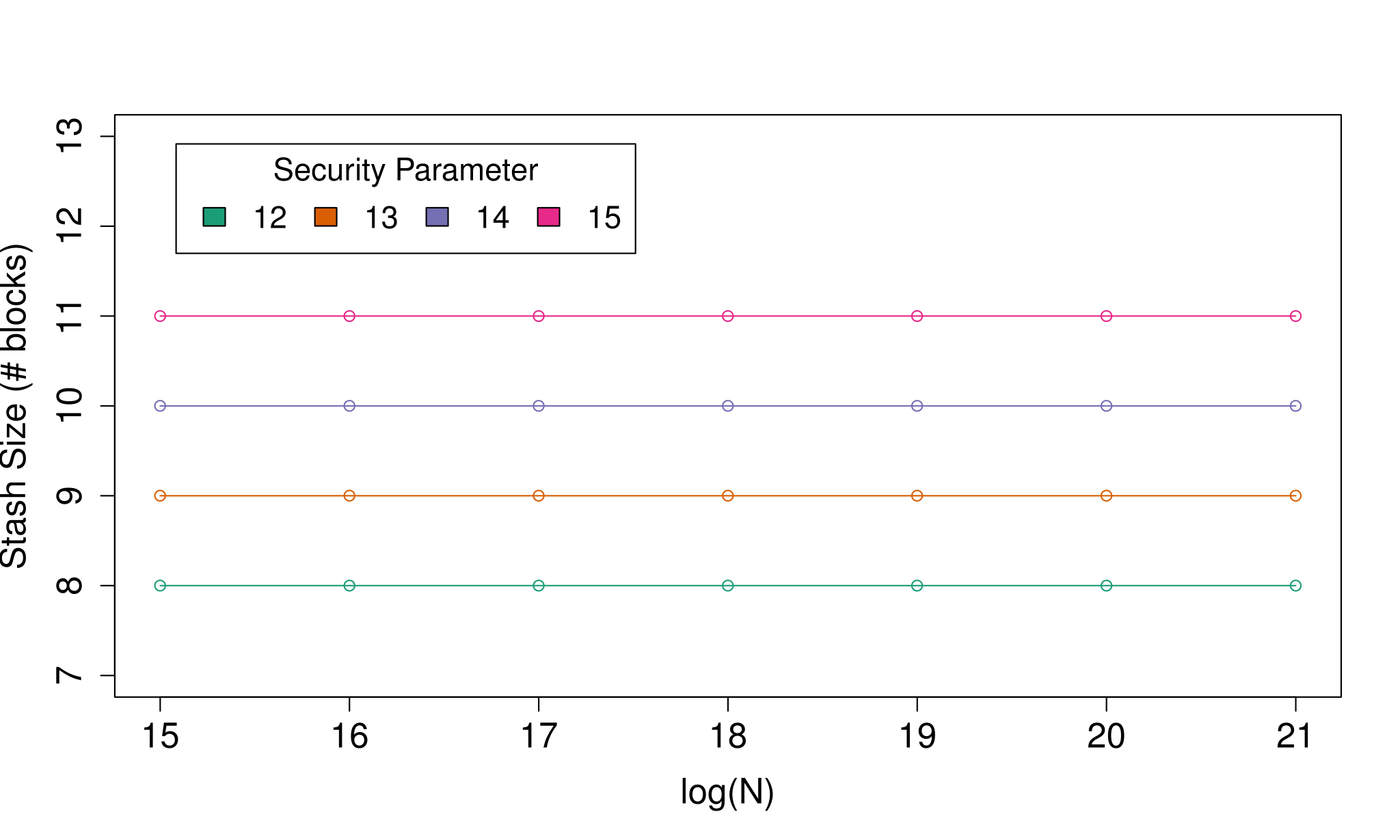}
\caption{The stash size to achieve failure probability less than $2^{-\lambda}$ does not depend on $N$ ($Z = 4$). Measured for a worst-case (in terms of stash size) access pattern. The stash size does not include the temporarily fetched path during \access.}
\label{fig:CacheSizes-VaryingN}
\end{figure}

\begin{figure}[t]
\centering
\begin{tabular}{|l|l|l|l|l|} \hline
\multirow{3}{*}{\bf Security Parameter ($\lambda$)} & \multicolumn{3}{c|}{\bf Bucket Size ($Z$)}\\ \cline{2-4}
& {\bf 4} & {\bf 5} & {\bf 6}\\ \cline{2-4}
& \multicolumn{3}{c|}{\bf Max Stash Size}\\ \hline
80 & 89 & 63 & 53 \\ \hline
128 & 147 & 105 & 89 \\ \hline
256 & 303 & 218 & 186 \\ \hline
\end{tabular}
\caption{{\bf Required max stash size for large security parameters.} Shows the maximum stash size required such that the probability of exceeding the stash size is less than $2^{-\lambda}$ for a worst-case (in terms of stash size) access pattern. Extrapolated based on empirical results for $\lambda \le 26$. The stash size does not include the temporarily fetched path during \access.}
\label{tab:StashSizesForNegligibleFailure}
\end{figure}

Our main theorem in Section \ref{sec:fail} shows the probability of exceeding stash
capacity decreases exponentially with the stash size, given that the bucket size $Z$ is large enough. 
This theorem is verified by experimental results as shown in Figure~\ref{fig:CacheSizes-VaryingN} and Figure~\ref{fig:CacheSizes-LambdaVsStashSize}.
In each experiment, the ORAM is initially empty. We first load $N$ blocks into ORAM and then access each block in a round-robin pattern.
I.e., the access sequence is $\{1, 2, \ldots, N, 1, 2, \ldots, N, 1, 2, \ldots\}$. 
In section~\ref{sec:fail}, we show that this is a worst-case access pattern in terms of stash occupancy for Path ORAM.
We simulate our Path ORAM for a single run for about 250 billion accesses after doing 1 billion accesses for warming-up the ORAM. 
It is well-known that if a stochastic process is regenerative (empirically verified to be the case for Path ORAM), the time average over a single run is equivalent to the ensemble average over multiple runs (see Chapter 5 of \cite{Harchol-BalterBook}).

Figure~\ref{fig:CacheSizes-LambdaVsStashSize} shows the minimum stash size to get a failure probability less than $2^{-\lambda}$ with $\lambda$ being the security parameter on the x-axis. In Figure~\ref{tab:StashSizesForNegligibleFailure}, we extrapolate those results for realistic values of $\lambda$. The experiments show that the required stash size grows linearly with the security parameter, which is in accordance with the Main Theorem in Section \ref{sec:fail} that the failure probability decreases exponentially with the stash size. Figure \ref{fig:CacheSizes-VaryingN} shows the required stash size for a low failure probability ($2^{-\lambda}$) does not depend on $N$. This shows Path ORAM has good scalability.

Though we can only prove the theorem for $Z \geq 5$, in practice, the stash capacity is not
exceeded with high probability when $Z=4$. $Z=3$ behaves relatively worse in terms of stash occupancy, and it is unclear how likely the stash capacity is exceeded when $Z=3$.

We only provide experimental results for small security parameters to show that the required stash size is $O(\lambda)$ and does not depend on $N$. 
Note that it is by definition infeasible to simulate for practically adopted security parameters (e.g., $\lambda = 128$), since if we can simulate a failure in any reasonable amount of time with such values, they would not be considered secure.

A similar empirical analysis of the stash size (but with the path included in the stash) was done by Maas \etal~\cite{phantom}.

\subsection{Bucket Load} \label{sec:BucketLoad}

\begin{figure*}[t]
	\centering
	\subfigure[Z=3]{\includegraphics[width=0.3\textwidth]{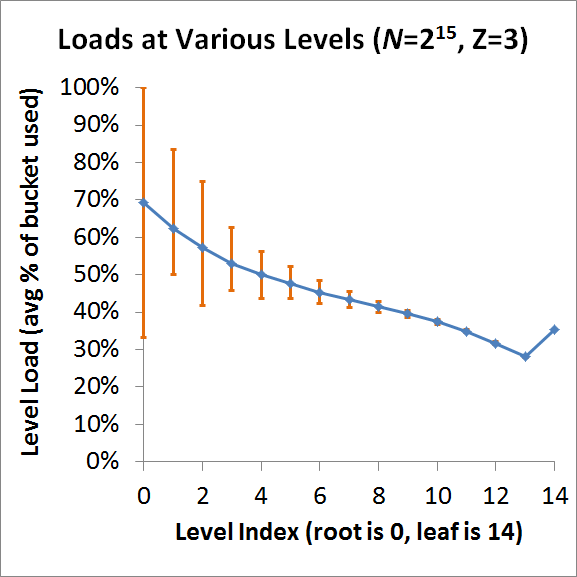}}
	\subfigure[Z=4]{\includegraphics[width=0.3\textwidth]{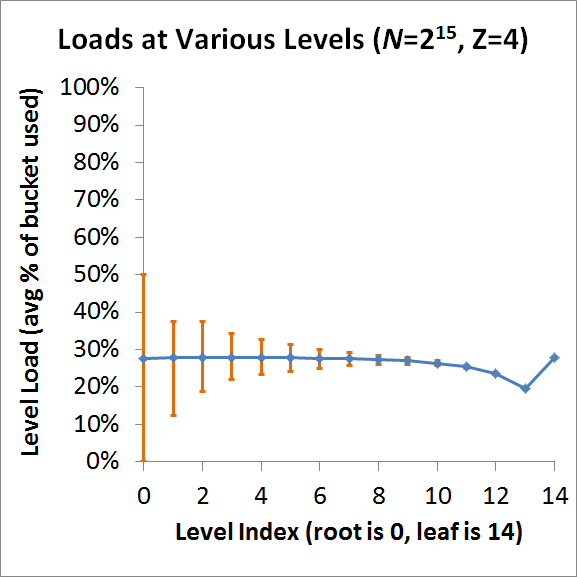}}
	\subfigure[Z=5]{\includegraphics[width=0.3\textwidth]{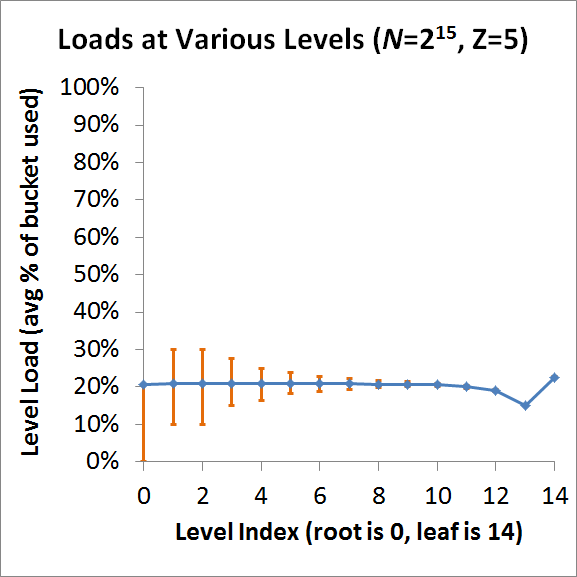}}
	\caption{Average bucket load of each level for different bucket sizes. The error bars represent the 1/4 and 3/4 quartiles.  Measured for a worst-case (in terms of stash size) access pattern.}
	\label{fig:LevelLoad}
\end{figure*}

Figure \ref{fig:LevelLoad} gives the bucket load per level for $Z \in \{3,4,5\}$. 
We prove in Section \ref{sec:fail} that for $Z \geq 5$, the expected usage of a subtree $T$ is close to the number of buckets in it. 
And Figure~\ref{fig:LevelLoad} shows this also holds for $4 \leq Z \leq 5$. 
For the levels close to the root, the expected bucket load is indeed 1 block (about 25\% for $Z=4$ and 20\% for $Z=5$). 
The fact that the root bucket is seldom full indicates the stash is empty after a path write-back most of the time.
Leaves have slightly heavier loads as blocks accumulate at the leaves of the tree. 
$Z=3$, however, exhibits a different distribution of bucket load (as mentioned in Section~\ref{sec:StashDistribution} and shown in Figure~\ref{fig:CacheSizes-LambdaVsStashSize}, $Z=3$ produces much larger stash sizes in practice).

\section{Conclusion}
Partly due to its simplicity, Path ORAM is the most practical ORAM scheme known-to-date under a small amount of client storage. We formally prove asymptotic bounds on Path ORAM, and show that its performance is competitive with or asymptotically better than the the best known construction (for a small amount of client storage), 
assuming reasonably large block sizes. We also present simulation results that confirm our theoretic bounds.

\section*{Acknowledgments}
Emil Stefanov was supported by the National Science Foundation Graduate Research Fellowship under Grant No. DGE-0946797 and by a DoD National Defense Science and Engineering Graduate Fellowship. 
Elaine Shi is supported by NSF under grant CNS-1314857.
Christopher Fletcher was supported by a National Science Foundation Graduate Research Fellowship, Grant No. 1122374 and by a DoD National Defense Science and Engineering Graduate Fellowship. This research was partially supported by the DARPA Clean-slate design of Resilient, Adaptive, Secure Hosts (CRASH) program under contract N66001-10-2-4089, and a
grant from the Amazon Web Services in Education program. Any opinions, findings, and conclusions or recommendations expressed in this material are those of the author(s) and do not necessarily reflect the views of the funding agencies.

We would like to thank Kai-Min Chung and Jonathan Katz for helpful discussions. We would like to thank Kai-Min Chung
for pointing out that our algorithm is information-theoretically (statistically) secure.

\nocite{GSORAM,oram01,oram02,oram03,GMOT11,oram05,oram06,oram07,oram08,oram09,GoldORAM,OsORAM,oram12,oram13,oramtrustedhardware01,oramtrustedhardware02,oramtrustedhardware03,ndss12,asiacrypt11}

\bibliographystyle{abbrv}
\bibliography{refs}

\end{document}